\documentclass[letterpaper, 10 pt, conference]{ieeeconf}  % Comment this line out
\IEEEoverridecommandlockouts                              % This command is only
\usepackage{graphics} % for pdf, bitmapped graphics files
\usepackage{epsfig} % for postscript graphics files
\usepackage{mathptmx} % assumes new font selection scheme installed
\usepackage{times} % assumes new font selection scheme installed
\usepackage{amsmath} % assumes amsmath package installed
\usepackage{amssymb}  % assumes amsmath package installed
\usepackage{xcolor}
\usepackage{tikz}

\usepackage{mathtools}

\title{\LARGE \bf
Sample-based observability of linear discrete-time systems
}

\author{Isabelle Krauss, Victor G. Lopez and Matthias A. Müller% <-this % stops a space
	\thanks{This work received funding from the European Research Council (ERC) under the European Union’s Horizon 2020 research and innovation programme (grant agreement No 948679).}% <-this % stops a space
	\thanks{I. Krauss, V. G. Lopez and M. A. Müller are with the  Leibniz University Hannover, Institute of Automatic Control, 30167 Hannover, Germany
		{\tt\small \{krauss,lopez,mueller\}@irt.uni-hannover.de}}%
}

\newif\iflong
\longtrue % switch to false in short version
\newcommand{\inLongVersion}[1]{\iflong #1\fi}

\newcommand\farbig[1]{\textcolor{black}{#1}}

\newcommand\copyrighttext{%
	\footnotesize \copyright 2022 IEEE. Personal use of this material is permitted. Permission from IEEE must be obtained for all other uses, in any current or future media, including reprinting/republishing this material for advertising or promotional purposes, creating new collective works, for resale or redistribution to servers or lists, or reuse of any copyrighted component of this work in other works.}
\newcommand\copyrightnotice{%
	\begin{tikzpicture}[remember picture,overlay]
		\node[anchor=south,yshift=10pt] at (current page.south) {\fbox{\parbox{\dimexpr\textwidth-\fboxsep-\fboxrule\relax}{\copyrighttext}}};
	\end{tikzpicture}%
}
\begin{document}
		\newtheorem{thm}{Theorem}%[section]
	\newtheorem{cor}{Corollary}
	\newtheorem{lem}{Lemma}

	\newtheorem{rem}{Remark}

	\newtheorem{defi}{Definition}
	\newtheorem{ass}{Assumption}
	\newtheorem{con}{Condition}

\maketitle
\thispagestyle{empty}
\pagestyle{empty}
\copyrightnotice

%%%%%%%%%%%%%%%%%%%%%%%%%%%%%%%%%%%%%%%%%%%%%%%%%%%%%%%%%%%%%%%%%%%%%%%%%%%%%%%%
\begin{abstract}
In this work, sample-based observability of linear discrete-time systems is studied. That is, we consider the case where the system output measurements are not available at every time instance.  It is shown that some discrete-time systems exhibit particular behaviors that lead to pathological sampling. Depending on the characteristics of the system, different sampling schemes are developed that allow the system state to be reconstructed.
\end{abstract}

%%%%%%%%%%%%%%%%%%%%%%%%%%%%%%%%%%%%%%%%%%%%%%%%%%%%%%%%%%%%%%%%%%%%%%%%%%%%%%%%
\section{INTRODUCTION}

In many control applications, only few and infrequent output measurements are available. Such cases can arise due to expensive  sensor placement, or the characteristics of the application make it impractical or difficult to measure the system output continuously or at every sampling instant. 
Consider, for example, the case of biomedical applications, where taking blood samples from a patient for testing cannot be performed too often. This is, e.g., an issue when determining certain hormone concentrations from blood samples in order to detect disorders of the hypothalamic–pituitary–thyroid axis \cite{b1} and to devise suitable medication dosages, compare, e.g., \cite{b2}. \farbig{Irregluar sampling sequences can also be generated due to event-triggered sampling of networked control systems to save computation and communication resources \cite{b10}}. To apply state-feedback control techniques for such systems, suitable state estimators need to recover the internal state by using only this limited output information. Sample-based observability is the area of research that studies the conditions that must be satisfied to ensure that the selected measurements are sufficient to reconstruct the system state.\par

Different conditions for sample-based observability have been studied for continuous-time linear systems.
The most basic instance corresponds to the case where the continuous-time system is discretized under uniform sampling and the sampling period is non-pathological. 
Sampling periods are said to be pathological, if the discretized system does not preserve the observability and controllability properties of the original system \cite{b3}. For every pair of eigenvalues $(\lambda_p,\lambda_q)$ of the system matrix, the sampling period $T$ needs to satisfy the following to rule out pathological situations 
\begin{align}
	\begin{aligned}
		&\text{if} \ \Re{(\lambda_p})= \Re{(\lambda_q}), \\ &\text{then} \ \Im{(\lambda_p)} \neq \Im{(\lambda_q)}+\frac{2k\pi}{T}, \ \forall k \in \mathbb{Z},
	\end{aligned}
	\label{eq:path_con}
\end{align} 
where $\Re{(\cdot)}$ and $\Im{(\cdot)}$ denote the real and imaginary parts, respectively \cite{b3}. 
In \cite{b4} a periodic sampling scheme was presented, that guarantees for every observable continuous-time system observability of the discretized system. Here, each period consists of a sequence of equidistant samples followed by one sample interval of different length. The average sampling period is now independent of the system parameters and can, thus, be chosen arbitrarily.
A slightly different sampling scheme was introduced in \cite{b5}. Here, a sequence of sampling instances is chosen nonequidistant. However, the sampling process is then periodically repeated.\par

More recent results in \cite{b6} and \cite{b7} allow completely irregular sampling of the continuous-time system. In these papers, a lower bound on the number of necessary samples taken at arbitrary time instances is derived to guarantee that the internal state can be reconstructed from the irregularly selected output measurements. To achieve this, an upper bound on the number of time instances where the system output can be zero for nonzero initial states inside the time period of interest is computed.\par 
In spite of these results for continuous-time systems, to the best of our knowledge, no results have been reported for sample-based observability of inherently discrete-time systems.
In this paper, we directly study conditions for sample-based observability of discrete-time systems. This addresses systems that are discrete by nature, as well as systems for which only discrete-time models are available. While some of the results we obtain are conceptually similar to those for continuous-time systems, other aspects are fundamentally different and inherent to the considered discrete-time setting. In particular, our contributions are as follows.
First, we investigate the conditions for discrete-time systems that lead to pathological sampling periods, analogous to the known conditions (\ref{eq:path_con}) for continuous-time systems. On the one hand, we show that this result directly allows for the development of a sampling scheme for arbitrary second-order systems. On the other hand, our analysis gives an insight into the difficulties that arise for higher-order systems.  
Then, we focus on designing sampling schemes for specific scenarios. For systems with only real eigenvalues, we derive a lower bound on the number of necessary samples taken at arbitrary time instances to guarantee sample-based observability. If the system has arbitrary eigenvalues and we are able to select specific samples to collect, then we show that a regular sampling scheme is necessarily non-pathological if the sampling period is properly selected. Finally, we discuss an extension to the regular sampling result to allow for specific irregular measurements.\par

The rest of the paper is organized as follows. In Section~\ref{sec:prob}, we formulate the sample-based observability problem. An analysis of pathological sampling in discrete-time systems, as well as our first sampling scheme results, are presented in Section \ref{sec:prop}. Section \ref{sec:schemes} explores different sampling schemes for discrete-time systems. The conclusions of the paper are given in Section \ref{sec:con}.

	\section{Problem formulation}
\label{sec:prob}
Consider a MISO (Multiple-Input Single-Output) linear time-invariant discrete-time system
\begin{align}
	&x(t+1)=Ax(t)+Bu(t) \label{eq:sysx}\\
	&y(t)=Cx(t)+Du(t)  \label{eq:sysy}
\end{align}
where $A \in \mathbb{R}^{n \times n}$, $B \in \mathbb{R}^{n \times m}$, $C \in \mathbb{R}^{1 \times n}$, $D \in \mathbb{R}^{1 \times m}$, $u(t) \in \mathbb{R}^m$ is the control input, $x(t) \in \mathbb{R}^n$ is the state and $y(t) \in \mathbb{R}$ is the system output.\\

\begin{defi}[Observability {\cite{b8}}]
	The system (\ref{eq:sysx})-(\ref{eq:sysy}) is said to be observable if for any unknown initial state $\xi$, there exists a finite $t>0$ such that the knowledge of the input $u$ and the output $y$ over $[0,t]$ suffices to determine uniquely the initial state $\xi$.\\
	\label{def:obs}
\end{defi}

Suppose that the well-known full rank condition of the observability matrix is satisfied and, thus, the system~\mbox{(\ref{eq:sysx})-(\ref{eq:sysy})} is observable \cite{b8}. Then, the following holds and, hence, every two initial states $\xi \neq \eta$ are distinguishable, i.e., 
\begin{align}
	\begin{pmatrix}
		\Delta y(0)\\
		\Delta y(1)\\
		\vdots\\
		\Delta y(n-1)
	\end{pmatrix}:=\begin{pmatrix}
		C\\
		CA\\
		\vdots\\
		CA^{n-1}
	\end{pmatrix}(\xi-\eta) \neq 0_n,
	\label{eq:rank_con}
\end{align}
where $0_{n} \in \mathbb{R}^{n}$ is a vector of zeros. If measurements are not available at $n$ consecutive samples, the initial state can always be recovered from observations of the output if and only if, for any $\xi$ and $\eta$, $\xi \neq \eta$,
\begin{align}
	\begin{pmatrix}
		\Delta y(t_1)\\
		\Delta y(t_2)\\
		\vdots\\
		\Delta y(t_l)
	\end{pmatrix}=\begin{pmatrix}
		CA^{t_1}\\
		CA^{t_2}\\
		\vdots\\
		CA^{t_l}
	\end{pmatrix}(\xi-\eta) \neq 0_n.
	\label{eq:rank_con}
\end{align} 
Therefore, we consider an observability rank condition depending on the set of sample instances ${t_1, t_2 \ldots, t_l}$.\\

\begin{defi}[Sample-based observability]
	The system in (\ref{eq:sysx})-(\ref{eq:sysy}) is sample-based observable if the sample-based observability matrix
	\label{def:obmat_s}
\end{defi}
\begin{align}
	\begin{pmatrix}CA^{t_1}\\CA^{t_2} \\ \vdots \\ CA^{t_l}\end{pmatrix}, \quad l\geq n,
	\label{eq:obmat_s}
\end{align}
has rank $n$.\par

In this paper, we investigate sampling schemes for the selection of the sample instances $t_1,\ldots ,t_l$, depending on the system dynamics such that the sample-based observability matrix (\ref{eq:obmat_s}) has full rank. Without loss of generality, the system will be considered in Jordan canonical form. Moreover, the following assumption is made. \\% on the system (\ref{eq:sysx})-(\ref{eq:sysy}).\par

\begin{ass}
	\label{ass:1a} $(A,C)$ is an observable pair.\\
\end{ass}

Additionally, we impose the following assumption on the system matrix $A$.\\
\begin{ass}
\label{ass:1b} All eigenvalues of the matrix $A$ are nonzero.\\
\end{ass}
 
To briefly illustrate the meaningfulness of Assumption \mbox{\ref{ass:1b}}, consider the case where matrix $A$ is given by \begin{align}
	A=	\begin{pmatrix}
		0&1\\
		\ & \ddots &\ddots\\
		\ & \ & \ddots & 1\\
		\ & \ & \ & 0\\
	\end{pmatrix}
\end{align}
This yields $A^k=0_{nn}$,  for all $k\geq n$, where $0_{nn} \in \mathbb{R}^{n \times n}$ is a matrix of zeros. Therefore, to satisfy the condition in Definition \ref{def:obmat_s} in this example, it is necessary to measure the first $n$ consecutive output samples, and no other sampling scheme can be obtained. In general, a system with a $p-$dimensional Jordan block of the eigenvalue zero requires measuring the first $p$ samples, independent of the other system characteristics and sampling scheme for later measurements.\par

\section{Pathological sampling sequences of discrete-time systems}
\label{sec:prop}
In this section, we provide conditions for sample-based observability of two-dimensional systems, and describe the properties present in some discrete-time systems that prevent this result from being generalized to higher-dimensional systems. To begin this analysis, we first study the conditions that result in pathological sampling.
\subsection{Pathological periods}
In the following, we take a closer look at some notable behavior of discrete-time systems, that is closely related to pathological sampling periods of the system  and, thus, is crucial for the development of possible sampling schemes. \par
Recall that we consider the system matrix $A$ in (\ref{eq:sysx}) in Jordan form, and denote the eigenvalue corresponding to the $i$-th Jordan block by $\lambda_i$. Then, for $h \in \mathbb{N}$, $A^h$ is a block diagonal matrix with each block being an upper triangular matrix with the same eigenvalue $\lambda_i^h$ on the diagonal. 
Notice that if $\lambda_p^h = \lambda_q^h$, for two eigenvalues $\lambda_p$, $\lambda_q$ of $A$ \farbig{with $p \neq q$} and some $h \in \mathbb{N}$, then $A^h$ has multiple blocks with the same eigenvalue. 
Moreover, this implies, in the case of single-output systems, that $(A^h,C)$ is not an observable pair. \inLongVersion{This can, e.g., be easily seen from the Hautus test for observability \cite{b85}, since the matrix $[C^\top, (\lambda_p^hI - A^h)^\top ]^\top$ has (at least) two rows of all zeros, and hence cannot have rank $n$ as required for observability.} The following lemma states conditions for this behavior to arise.\\

\begin{lem}
	Consider two eigenvalues $\lambda_p,\lambda_q$ that belong to two different Jordan blocks of $A$. Then, $\lambda_p^h = \lambda_q^h$ if and only if $|\lambda_p|=|\lambda_q|$ and there exists a positive integer $h$ that satisfies 
	\begin{align}
	\frac{\pi}{\phi_q-\phi_p}=\frac{h}{2(k_q-k_p)}
	\label{eq:lemh}
	\end{align}
	\farbig{for some $k_p, k_q \in\{0,1,\ldots,h-1\}$, $k_p\neq k_q$, where $\phi_i$ is the phase of the complex number $\lambda_i$,  $i=p,q$.}
\label{lem:h}
\end{lem}

\begin{proof}
Notice that the $h$-th power of a complex number $\lambda_p$ can be expressed as
\begin{align}
	\lambda_p^h=|\lambda_p|^he^{h\phi_p j}=|\lambda_p|^h(\cos(h\phi_p)+j \sin(h \phi_p)).
	\label{eq:powerh}
\end{align}
To have $\lambda_p^h=\lambda_q^h$, both $\lambda_p$ and $\lambda_q$ must be an $h$-th root of some $\gamma=|\gamma|e^{\theta j} \in \mathbb{C}$
\begin{align}
	\sqrt[h]{\gamma}=|\gamma|^\frac{1}{h}e^{\frac{\theta}{h} j}=|\gamma|^\frac{1}{h}(\cos(\frac{\theta}{h})+j \sin(\frac{\theta}{h})).
	\label{eq:root_gamma}
\end{align}
Therefore, \begin{align}
	\phi_p=\frac{\theta+2\pi k_p}{h},\\
	\phi_q=\frac{\theta+2\pi k_q}{h},
\end{align}
for some $k_p, k_q=0,1,\ldots,h-1$, $k_p\neq k_q$. Hence, $\lambda_p$ and $\lambda_q$ are both an $h$-th root of $\gamma$ if and only if $|\lambda_p|=|\lambda_q|$ and~(\ref{eq:lemh}) holds.
\end{proof}
\begin{rem}
	Notice that if $\lambda_p=-\lambda_q$, then (\ref{eq:lemh}) is satisfied for every even $h$. Furthermore, if $(\lambda_p,\lambda_q)$ is a complex conjugate pair, then $\phi_p=-\phi_q$ and $\gamma$ in the proof of Lemma~\ref{lem:h} is a real number. Thus, it follows directly from (\ref{eq:powerh}) that \mbox{$h\phi_p=\pi k$}, for some $k \in \{\pm1,\ldots, \mbox{$\pm(h-1)\}$}$. Hence, for a complex conjugate pair, (\ref{eq:lemh}) can be replaced by
	\begin{align}
		\pi / \phi_p=h/k. 
	\end{align}
\end{rem}
\begin{rem}
	Consider a pair of eigenvalues $(\lambda_p,\lambda_q)$ that satisfies the conditions in Lemma~\ref{lem:h}. Then, notice that all values of $h$ for which (\ref{eq:lemh}) holds are an integer multiple of some minimum value $\bar{h}$.\\
	\label{rem:h}
\end{rem}

Now, Lemma~\ref{lem:path} on pathological measurement sequences can be formulated.\\ 

\begin{lem}
	Consider system (\ref{eq:sysx})-(\ref{eq:sysy}). %and let Assumptions \ref{ass:1a} and \ref{ass:1b} hold. 
	If the conditions in Lemma~\ref{lem:h} hold, then any measurement sequence of the form
		\begin{align}
		t_i=r_i h,  
		\label{eq:path_seq}
	\end{align}
with arbitrary $r_i \in \mathbb{N}, i=1,\ldots,l$, and $l \in \mathbb{N}$,
 results in a sample-based observability matrix (\ref{eq:obmat_s}) with rank less than~$n$.\label{lem:path}	
\end{lem}
\begin{proof}
If the conditions in Lemma~\ref{lem:h} are satisfied and, thus, $\lambda_p^h=\lambda_q^h$, then  $A^h$ has multiple blocks with the same eigenvalue, and hence $(A^h,C)$ is not an observable pair as discussed above Lemma~\ref{lem:h}. Therefore the sample-based observability matrix (\ref{eq:obmat_s}) with $t_i$ as in (\ref{eq:path_seq}) has always a rank less than $n$.
\end{proof}

\subsection{Sampling scheme for second-order systems}
From the preceding analysis of pathological sampling sequences for discrete-time systems, conditions on the sampling for second order systems to guarantee sample-based observability can be straightforwardly concluded.\\
\begin{lem} 
	For a system (\ref{eq:sysx})-(\ref{eq:sysy}) with $n=2$ satisfying \text{Assumptions \ref{ass:1a} and \ref{ass:1b}}, the sample-based observability matrix~(\ref{eq:obmat_s}) has full rank for any two measurement instances, if the conditions in Lemma  \ref{lem:h} are not satisfied. 
	If the conditions in Lemma~\ref{lem:h} hold, then sample-based observability is guaranteed by measuring $N_s$ arbitrary samples inside any time interval $[t,t+T-1]$ with $t\in \mathbb{N}_0, T \in \mathbb{N}$, where
	\begin{align}
		N_s\geq 1+\frac{T}{h}
		\label{eq:Ns}
	\end{align}
	\label{lem:2d}
and $h$ assumes the minimum value $\bar{h}$ as in Remark \ref{rem:h}. 
\end{lem}
\begin{proof}
	Recall that we consider $A$ in Jordan form.
	Since $(A,C)$ is observable by Assumption \ref{ass:1a}, if $\lambda_1 = \lambda_2$, then $A$ is a Jordan block of dimension 2. This follows since otherwise, 
	\begin{align}
		\text{rank}\begin{pmatrix}C \\ \lambda I -A\end{pmatrix} <2
	\end{align}
	 for any $C$, for a single-output system, contradicting observability. 
	 Hence, in case that $\lambda_1=\lambda_2$, $A^h \neq \alpha I$ for any $\alpha\in \mathbb{R}$ (including $\alpha=0$ due to Assumption \ref{ass:1b}) and any $h\in \mathbb{N}$. Together with the fact that $C\neq 0$ by Assumption $\ref{ass:1a}$, this implies $CA^h \neq \alpha C$ for any $\alpha\in \mathbb{R}$ and any $h\in \mathbb{N}$ and, thus, that (\ref{eq:obmat_s}) has full rank for any two samples.
	In the case that $ \lambda_1 \neq \lambda_2$, $A$ is a diagonal matrix. If $\lambda_1^h \neq \lambda_2^h$ for all $h\in \mathbb{N}$, then again $CA^h \neq \alpha C$, showing that (\ref{eq:obmat_s}) has full rank for any two samples. 
	If $\lambda_1^h = \lambda_2^h$, let $h$ take the minimum value $\bar{h}$ as in Remark \ref{rem:h}. Without loss of generality, assume the first sample is selected at $t=0$. Then, out of $T$ samples, there exist up to $\frac{T}{h}$ time instances $t\neq 0$ that result in 
	\begin{align}
		C=\alpha_t CA^t, \ \alpha_t \in \mathbb{R}, \ t \in \mathbb{N}.
		\label{eq:pr2d}
	\end{align}
	Hence, taking $N_s$ samples as in (\ref{eq:Ns}) guarantees that there exists at least one time instance, for which (\ref{eq:pr2d}) does not hold.\end{proof}

\subsection{Higher-order systems}	
\label{subsec:hos}
Lemma~\ref{lem:2d} cannot be extended straightforwardly to higher-order systems. Finding a lower bound for randomly taken measurements to ensure sample-based observability for general linear systems is a challenging task. In the following, we illustrate with an example some of the difficulties that arise for higher order systems.\par
Consider the case where $A^h$ has more than two blocks with the same eigenvalue and, hence, pathological sampling situations are more likely.
In particular, assume the worst-case scenario in which  $A^n$ has $n$ blocks with the same eigenvalue, i.e., 
\begin{equation}
	\begin{aligned}
	&\lambda_1^n=\lambda_2^n=\ldots=\lambda_n^n, \quad \lambda_p \neq \lambda_q, \\
	&  \farbig{\forall  p,q \in \{1, \ldots ,n\}, \ p \neq q. }
	\label{eq:eigs}
	\end{aligned}
\end{equation}
In Appendix \ref{App:B}, we provide a numerical example for which this is the case. Then, the set of time instances 
\begin{equation}
	\begin{aligned}
	&K:=\{t, t+r_1 n, t+r_2 n ,\ldots,t+r_{n-1} n \}, \\ 
	&\forall t \in \mathbb{N}_0, \ \forall r_i \in \mathbb{N}, \ i=1,\ldots,n-1,
	\end{aligned}
\end{equation}
generates a sample-based observability matrix (\ref{eq:obmat_s}) with only rank one, since $CA^t=\alpha_i CA^{t+r_in}$, 
for some $\alpha_i \in \mathbb{R}$. It can be shown that a bound on arbitrary samples in any interval $[t,t+T-1]$ to guarantee sample-based observability in such a case must satisfy
\begin{align}
	N_s\geq 1+\frac{n-1}{n}T,
	\label{eq:boundn}
\end{align}
as follows from the example in Appendix \ref{App:B}.
However, when collecting at least $1+T(n-1)/n$ out of $T$ samples, this includes at least one set of $n$ consecutive samples. This means that for a system with the characteristics in (\ref{eq:eigs}), no meaningful scheme for measuring arbitrary samples can be given.
As shown in the following lemma, to achieve sample-based observability without taking $n$ consecutive samples, measurements at specific time instances are required.  \\

\begin{lem}
	Consider system (\ref{eq:sysx})-(\ref{eq:sysy}) and let Assumptions \ref{ass:1a} and \ref{ass:1b} hold. Moreover, let the eigenvalues of matrix $A$ make (\ref{eq:eigs}) hold. Then, a sampling scheme leads to sample-based observability if and only if it contains a set of samples of the form
	%(\ref{eq:set_ex})
	\begin{equation}
		\begin{aligned}
			K=\{t,t+r_1 n+1,t+r_2 n+2, \ldots,t+r_{n-1} n+n-1\},
		\end{aligned}
	\label{eq:set_ex}
	\end{equation}
where $t,r_i \in \mathbb{N}_0, \ i=1,...,n-1$.\\
 \label{lem:r1ex}
\end{lem}

Proof: see Appendix \ref{App:A}.\\ 
%Proof: see Appendix \ref{App:A}.\\

In this section, conditions for pathological sampling sequences were derived, and it was observed that obtaining a (meaningful) bound on arbitrary samples in order to guarantee sample-based observability is not possible in general.  Nevertheless, we provided a first insight on possible sampling schemes for specific systems. In the following section, we will develop further sampling schemes for more general cases. 

\section{Sampling schemes for higher order systems}
\label{sec:schemes}
In the preceding section, we stated a sample-based observability result  for second-order systems. Now, we will focus on systems with dimension $n\geq 3$. First, the case for systems with only real eigenvalues will be studied, followed by our analysis of systems with both complex and real eigenvalues. 
\subsection{Systems with only real eigenvalues}\label{AA}
%Assumption: no Jordan blocks with eigenvalues of the same magnitude.\\
In this subsection, we consider linear discrete-time systems with exclusively real eigenvalues. A  lower bound on the number of arbitrary measurements is derived such that sample-based observability is guaranteed.\\

\begin{thm}
	Consider system (\ref{eq:sysx})-(\ref{eq:sysy}) and let Assumptions~\ref{ass:1a}~and~\ref{ass:1b} hold. Moreover, let $A$ have only real eigenvalues and let $(A^2,C)$ be an observable pair.  Then, any arbitrary $2n-1$ measurement instances $t_1, \ldots ,t_{2n-1}$, satisfy the sample-based observability condition in Definition \ref{def:obmat_s}.
	\label{thm:real}
\end{thm}
\begin{proof}
Consider the sample-based observability matrix for $n$ samples \begin{align}
	\begin{bmatrix}(CA^{t_1})^\top&(CA^{t_2})^\top &\ldots & (CA^{t_n})^\top\end{bmatrix}^\top.
	\label{eq:obmat_real}
	\end{align}
If (\ref{eq:obmat_real}) has \mbox{rank $<n$}, then $CA^{t_n}$ can be expressed as a linear combination of the first $n-1$ rows 
\begin{align}
	CA^{t_n}+\sum_{i=0}^{n-1}\alpha_iCA^{t_i}=0_n^\top, \ \alpha_i \in \mathbb{R},
	\label{eq:lin_c} 
\end{align}
where $C=(c_1, \ldots ,c_n)$. Notice that $A$ is in Jordan form, and hence $c_k\neq0$, when in the $k$-th column of $A$ a new Jordan block begins, since $(C,A)$ is an observable pair.
Thus, for (\ref{eq:lin_c}) to hold, each eigenvalue $\lambda_j$ %$\lambda_1, \ldots, \lambda_n$ of $A$ 
has to be a solution to
\begin{align}
	\lambda^{t_n}+\sum_{i=1}^{n-1}\alpha_i \lambda^{t_i}=0.
	\label{eq:pol_real}
\end{align}
In case of a $p$-dimensional Jordan block, the associated eigenvalue is a root of multiplicity $p$ of (\ref{eq:pol_real}). This is because this eigenvalue is also a root of all derivatives of (\ref{eq:pol_real}) up to the $p$-th derivative. For clarity, we illustrate this by considering, e.g.,  a third-order system with $A$ consisting of two Jordan blocks. Then, (\ref{eq:obmat_real}) can be written as  
\begin{align}
	\begin{pmatrix}
		c_1 \lambda_1^{t_1}&c_2 \lambda_2^{t_1}&c_2 t_1 \lambda_{2}^{t_1-1}+c_{3} \lambda_{2}^{t_1}\\
		c_1 \lambda_1^{t_2}&c_2 \lambda_2^{t_2}&c_2 t_2 \lambda_{2}^{t_2-1}+c_{3} \lambda_{2}^{t_2}\\
		c_1 \lambda_1^{t_3}&c_2 \lambda_2^{t_3}&c_2 t_3 \lambda_{2}^{t_3-1}+c_{3} \lambda_{2}^{t_3}
	\end{pmatrix}.
\label{eq:mat_3d}
\end{align}
From the first two columns of (\ref{eq:mat_3d}) it follows that, for (\ref{eq:lin_c}) to hold, $\lambda_1$ and $\lambda_2$ must both satisfy (\ref{eq:pol_real}). From the third column it follows, furthermore, that $\lambda_2$ is not only a root of  (\ref{eq:pol_real}) but a root of its derivative as well. Hence, $\lambda_2$ is a double root of (\ref{eq:pol_real}).  This third-order example can be straightforwardly extended to any $p$-dimensional Jordan block.
Now, we argue that the conditions in Theorem~\ref{thm:real} are such that for a specific set of $n$ out of the considered $2n-1$ samples, (\ref{eq:pol_real}) cannot have $n$ nonzero real solutions, contradicting (\ref{eq:lin_c}).
The polynomial in (\ref{eq:pol_real}) consists of $n$ terms and, hence, a maximum of $n-1$ sign changes in the coefficients can occur. Applying Descartes' rule of signs (see e.g. \cite{b9}), both the number of possible positive real solutions and the number of possible negative real solutions is upper bounded by $n-1$. 
Now, consider the case that all exponents of the polynomial (\ref{eq:pol_real}) are even. Then, (\ref{eq:pol_real}) has real solutions of the form $\pm\lambda_j$, $1\leq j<n$. Since $(A^2,C)$ being an observable pair rules out the case where $A$ has eigenvalues of the same magnitude but with different signs, (\ref{eq:pol_real}) with only even exponents cannot hold for all $n$ real nonzero eigenvalues of $A$. Thus we conclude that,  if all $t_i$ are even, the matrix in (\ref{eq:obmat_real}) has rank $n$.
Therefore, sample-based observability is achieved if at least $n$ of the measured samples correspond to even measurement instances $t_i$. Similarly, it is also sufficient to take $n$ odd instances. Finally, taking any $2n-1$ samples guarantees either $n$ even or $n$ odd measurement instances $t_i$, completing the proof. 
\end{proof}

From the presented proof, the following corollary for the case of only positive eigenvalues can be straightforwardly concluded.\\

\begin{cor}
	Let Assumptions \ref{ass:1a} and \ref{ass:1b} hold. If the system matrix $A$ has only positive real eigenvalues, the collection of any  $n$ samples results in rank $n$ of the sample-based observability matrix (\ref{eq:obmat_s}).\\
	\label{cor}
\end{cor}
\begin{rem} \label{rem} As shown in the preceding proof, the assumption of $(A^2,C)$ being an observable pair is required in Theorem~\ref{thm:real} to state that any $2n-1$ samples, regardless of the time of measurement, guarantee sample-based observability.
If the assumption of $(A^2,C)$ being an observable pair is not satisfied, only a time interval-dependent bound similar to Lemma~\ref{lem:2d} can be derived, i.e., in any interval $[t,t+T-1]$
	\begin{align}
		N_s\geq \frac{N_p+T}{2}, \ T \geq 2n
		\label{eq:boundA2}
	\end{align}
arbitrary samples  are required, where $N_p$ is the number of eigenvalues for which the conditions in Lemma~\ref{lem:h} hold for $h=2$. We show the bound in (\ref{eq:boundA2}) in Appendix \ref{App:C}.
\end{rem}
\subsection{Regular sampling scheme}
In the following, a regular \farbig{(periodic)} sampling scheme will be derived. This scheme requires regularly taken measurements, providing less freedom in the selection of samples.
 However, different from the conditions in Theorem~\ref{thm:real}, this sampling scheme requires measuring only $n$ samples, and is applicable also for systems with complex eigenvalues.\par

In Lemma~\ref{lem:path}, a condition for pathological measurement sequences due to the pathological sampling period $h$ for discrete-time systems was presented. Now, we will show that every regular sampling period other than $h$ is non-pathological. \\

\begin{thm}
Consider system (\ref{eq:sysx})-(\ref{eq:sysy}) and let Assumptions \ref{ass:1a} and \ref{ass:1b} hold.
\farbig{The set $\{t_1,t_1+\bar{t},t_1+ 2\bar{t}, \ldots, t_1+(n-1)\bar{t}\}$ consisting of $n$ equidistant time instances with $t_1 \in \mathbb{N}_0, \ \bar{t} \in \mathbb{N}$, guarantees that the sample-based observability condition in Definition (\ref{def:obmat_s}) holds, if  $\bar{t} \neq h$ with $h$ being any pathological sampling period as in Lemma~\ref{lem:h}.} \label{thm:reg}
\end{thm}
\begin{proof}
	The proof is performed by contradiction. Without loss of generality assume $t_1=0$. Suppose that the matrix
\begin{align}
	\begin{bmatrix} C^\top, (CA^{\bar{t}})^\top, (CA^{2\bar{t}})^\top, \ldots, (CA^{(n-1)\bar{t}})^\top \end{bmatrix}^\top
%	\begin{pmatrix} C \\ CA^{\bar{t}}%\\CA^{2\bar{t}}
	%	\\ \vdots \\CA^{(n-1)\bar{t}}\end{pmatrix} 
	\label{eq:obmat_reg}
\end{align}
is not full rank. Then, $CA^{(n-1)\bar{t}}$ is a linear combination of the first $n-1$ rows. \farbig{Hence, applying the same arguments as in the proof of Theorem~\ref{thm:real}, it can be seen that each eigenvalue $\lambda_j$ is a solution of multiplicity $p_j$ to
%the eigenvalues $\lambda_j$ of $A$ with an algebraic multiplicity $p_j$ are solutions  of  multiplicity $p_j$ to
\begin{align} 
	\lambda^{(n-1)\bar{t}}+\sum_{i=0}^{n-2}\alpha_i \lambda^{i\bar{t}}=0, \  \alpha_i\in \mathbb{R},
	\label{eq:pol_reg}
\end{align}
where $p_j$ is the algebraic multiplicity of $\lambda_j$. %\newpage
%\begin{flushleft}Now subsitute
Substituting $\lambda^{\bar{t}}$ by $\gamma$ in (\ref{eq:pol_reg}) results in 
\begin{align}
	\gamma^{(n-1)}+\sum_{i=0}^{n-2}\alpha_i \gamma^{i}=0.
	\label{eq:pol_reg2}
\end{align}
It can be shown that if $\lambda_j$ is a root of multiplicity $p_j$ of (\ref{eq:pol_reg}), then $\gamma_j=\lambda_j^{\bar{t}}$ is a root of the same multiplicity $p_j$ of  (\ref{eq:pol_reg2}). 
%Notice that if $\lambda_j$ is a root of multiplicity $p_j$ of (\ref{eq:pol_reg}), then $\gamma_j=\lambda_j^{\bar{t}}$ is a root of the same multiplicity $p_j$ of  (\ref{eq:pol_reg2}). 
Furthermore, due to the assumption that $\bar{t} \neq h$, it always holds that \mbox{$\lambda_p^{\bar{t}}\neq \lambda_q^{\bar{t}}$} for all pairs of eigenvalues $(\lambda_p,\lambda_q)$ with $\lambda_p\neq \lambda_q$. Finally, since (\ref{eq:pol_reg2}) is a polynomial of order $n-1$, and hence has only $n-1$ solutions, we can conclude that the above implies that (\ref{eq:pol_reg}) does not hold for all $n$ eigenvalues.
Therefore, (\ref{eq:obmat_reg}) has full rank.}
\end{proof}

\subsection{Irregular sampling scheme}
In the following, we will take a closer look at the possibility of modifying the regular sampling scheme in Theorem~\ref{thm:reg}  to allow some irregularity of the measurement instances.
First, a system of order three will be considered and subsequently the extension to systems with dimension $n>3$ is discussed.\par
\subsubsection{Third-order system}
For a third-order system such that
\begin{align}
C \neq \alpha CA^{t}, \ \forall t \in\mathbb{N}, \ \forall \alpha \in \mathbb{R},
\label{eq:CCA}
\end{align}
 we can derive a sampling scheme consisting of only four samples with mild conditions on the time instances of the measurements to guarantee full rank of the sample-based observability matrix. With  (\ref{eq:CCA}) being satisfied, we exclude systems for which the sample-based observability matrix can have rank one as in the case described in Subsection~\ref{subsec:hos}.\\

\begin{thm}
	Consider system (\ref{eq:sysx})-(\ref{eq:sysy}) and let Assumptions \ref{ass:1a} and \ref{ass:1b} hold. \farbig{Moreover suppose (\ref{eq:CCA}) is satisfied.
		Consider any two samples $t_1$ and $t_2$ and select the two additional samples $t_3=t_1+\Delta$ and $t_4=t_2+\Delta$ with $\Delta \in \mathbb{N}$. Then, the sample-based observability matrix (\ref{eq:obmat_s}) has rank $n$, if   $\Delta\neq h$ with $h$ being any pathological sampling period as in Lemma~\ref{lem:h}.} \label{thm:3d}
\end{thm}

\begin{proof}
	Assume that $\begin{bmatrix} (CA^{t_1})^\top, (CA^{t_2})^\top,(CA^{t_3})^\top \end{bmatrix}^\top$ is not full rank and, hence, there exist scalars $\alpha_1, \alpha_2 \in \mathbb{R}$ such that
	\begin{align}
		CA^{t_2}=\alpha_1 CA^{t_1}+\alpha_2  CA^{t_3},
		\label{eq:t2}
	\end{align} 
	with $t_3=t_1+\Delta$. 
	Then, we claim that there exist no scalars $\beta_1,\beta_2 \in \mathbb{R}$ such that 
	\begin{align} CA^{t_4}=\beta_1 CA^{t_1}+\beta_2  CA^{t_3},
		\label{eq:t4} 
	\end{align}
	with $t_4=t_2+\Delta$. 
	We show this claim by contradiction. To do so, assume that there exist $\beta_1$ and $\beta_2$ satisfying (\ref{eq:t4}). Notice that all $\alpha_1, \alpha_2,\beta_1, \beta_2\neq 0$ since otherwise it would contradict (\ref{eq:CCA}). Then
	\begin{align} 
		&CA^{t_4}=CA^{t_2+\Delta}\stackrel{(\ref{eq:t4})}{=}\beta_1 CA^{t_1}+\beta_2  CA^{t_3}\\
		&\phantom{CA^{t_4}=CA^{t_2+\Delta}}
		\stackrel{(\ref{eq:t2})}{=}\alpha_1 CA^{t_1+\Delta}+\alpha_2  CA^{t_1+\Delta+\Delta} \\
		&\Rightarrow CA^{t_1+2\Delta}=\frac{\beta_1}{\alpha_2} CA^{t_1}+\frac{\beta_2-\alpha_1}{\alpha_2} CA^{t_1+\Delta}.
	\end{align}
	However, this implies
	\begin{align}
		\text{rank} \begin{pmatrix} CA^{t_1}\\CA^{t_1+\Delta}\\CA^{t_1+2\Delta}\end{pmatrix}=2
	\end{align}
	contradicting Theorem~\ref{thm:reg}, which completes the proof.
\end{proof}

\subsubsection{Discussion on an extension to higher-order systems}
For discussing an extension of Theorem~\ref{thm:3d} to higher-order systems, first the following lemma is required. \\
\begin{lem}
	Consider system (\ref{eq:sysx})-(\ref{eq:sysy}) and let Assumptions \ref{ass:1a} and \ref{ass:1b} hold. \farbig{Suppose for the set $\{t_1,t_2 \ldots t_n\}$  of arbitrary samples the sample-based observability matrix (\ref{eq:obmat_s}) has a rank lower than $n$. Then, the additional samples $\{t_1+\Delta,t_2+\Delta, \ldots, t_n+\Delta\}$ will increase the rank of the sample-based observability matrix for all $\Delta\neq  h$ with $h$ being any pathological sampling period as in Lemma~\ref{lem:h}.}\\
	\label{lem:rank_in}
\end{lem}

\begin{proof} If for the set  $\{t_1,t_2 \ldots t_n\}$  the sample-based observability matrix has not full rank, then
\begin{align}
	CA^{t_n}=\sum_{i=1}^{n-1}\alpha_iCA^{t_i}
\end{align}
for some scalars $\alpha_i\in \mathbb{R}$. % not all zero.
To argue by contradiction, suppose that
\begin{align}
	CA^{t_j+\Delta}=\sum_{i=1}^{n-1}\gamma_{i,j}CA^{t_i}, \quad j=1, \ldots ,n
	\label{eq:sum_delta}
\end{align}
for some scalars $\gamma_{i,j} \in \mathbb{R}$. Hence, the rank of the sample-based observability matrix will not be increased by the additional set of samples.
Applying (\ref{eq:sum_delta}) $r$ times, it follows that for each $r=1,\ldots,n-1$ we have
\begin{align}
	CA^{t_n+r\Delta}=\sum_{i=1}^{n-1}\gamma_{i,n}CA^{t_i+(r-1)\Delta}=\cdots=\sum_{i=1}^{n-1}\beta_{i,r}CA^{t_i}
\end{align}
for some $\beta_{i,r}\in \mathbb{R}$.
Consequently this leads to 
\begin{align}
	\text{rank} \	\begin{pmatrix} CA^{t_n} \\ CA^{t_n+\Delta}\\CA^{t_n+2\Delta}\\ \vdots \\CA^{t_n+(n-1)\Delta}\end{pmatrix} <n.
	\label{eq:obmat_delta}
\end{align}
Since the set of measurement instances $\{t_n,t_n+\Delta, \ldots,t_n+(n-1)\Delta\}$ shows regularity and $\Delta\neq h$, inequality (\ref{eq:obmat_delta}) contradicts Theorem~\ref{thm:reg}.
\end{proof}

Using Lemma~\ref{lem:rank_in} to extend Theorem~\ref{thm:3d} to higher order systems yields a set $K_{n-1}$ of sufficient samples. $K_{n-1}$ is constructed in the following manner
\begin{equation}
	\begin{matrix*}[l]
		&K_1=\{t_1,t_2\},\\
		&K_2=\{K_1,K_1+\Delta_1\}, \\
		&K_3=\{K_2,K_2+\Delta_2\}, \\
		&\ \vdots\\
		&K_{n-1}=\{K_{n-2},K_{n-2}+\Delta_{n-2}\}.
	\end{matrix*}
\label{eq:Ki}
\end{equation}
Here, $K_i+\Delta$ is defined as the set of time instances resulting from adding a scalar $\Delta$ to each element of $K_i$. By Lemma~\ref{lem:rank_in}, each of the sets $K_i$ in (\ref{eq:Ki}) guarantees the addition of at least 1 to the rank of matrix (\ref{eq:obmat_s}), when compared to the set $K_{i-1}$, in case the rank is still less than $n$. However, this sampling scheme may not be convenient for high order systems, since the number of necessary samples  increases exponentially with the system order. In particular, for an \mbox{$n^{th}$-order} system, $2^{n-1}$ samples are required in the set $K_{n-1}$. Hence, compared to the regular sampling scheme from Theorem~\ref{thm:reg}, using irregular sampling (i.e., $\Delta_i \neq \Delta_j$) comes at the price of a (significantly) higher bound on the required maximum number of samples in order to guarantee sample-based observability.

\section{Conclusion}
\label{sec:con}
In this paper, we investigated the observability of linear discrete-time systems for the case that measurements are not available at every time instance. \par
First, pathological sampling sequences were analyzed. Second, we showed that for systems with only real eigenvalues, under certain conditions an arbitrary choice of \mbox{$2n-1$} (irregular) sampling instances is sufficient for sample-based observability.
Furthermore, building on the results on pathological sampling sequences, it was shown that for every sampling period other than the pathological sampling period $h$ as in Lemma~\ref{lem:h}, a regular sampling scheme leads to sample-based observability for general discrete-time systems (with real and/or complex eigenvalues). It was also shown how to relax the strict regularity of the sampling scheme while still guaranteeing sample-based observability; however, this might only be convenient for lower-order systems. \\
\indent Relaxing the conditions for an irregular sampling scheme for high order systems with general real and/or complex eigenvalues is an interesting subject of future research.

\appendix
\subsection{Proof of Lemma~\ref{lem:r1ex}}
\label{App:A}
\begin{proof}
	 If the set $K$ results in a rank-deficient sample-based observability matrix, then there exist $\alpha_i\in\mathbb{R}, \ i=0,...,n-1$, not all zero, such that
\begin{align}
	\sum_{i=0}^{n-1} \alpha_i CA^{t+r_in+i}=0_n^\top, \quad r_0=0.
	\label{eq:sumCA}
\end{align}
Multiplying (\ref{eq:sumCA}) from the right by $A^{-t}$ (which is well defined due to Assumption \ref{ass:1b}) yields
\begin{align}
	\sum_{i=0}^{n-1} \alpha_iCA^{r_in+i}=0_n^\top.
\end{align}
Substituting $A^n=\gamma I$ for some $\gamma=\lambda_1^n\neq 0$ due to (\ref{eq:eigs}) then leads to
\begin{align}
	\sum_{i=0}^{n-1} \alpha_i\gamma^{r_i}CA^{i}=0_{n}^\top.
	\label{eq:sumA}
\end{align}
However, (\ref{eq:sumA}) contradicts observability of the system, completing the proof of sufficiency of Lemma~\ref{lem:r1ex}.
Now, we show that selecting the samples according to (\ref{eq:set_ex}) is a necessary condition for sample-based observability.
Consider any set of samples such that the sample-based observability matrix (\ref{eq:obmat_s}) has rank $n$. Then, for any $\alpha_i$, $i=1,\ldots,l$, which are not all zero
\begin{align}
	\sum_{i=1}^{l} \alpha_iCA^{t_i}\neq0_{n}^\top.
		\label{eq:sumA2}
\end{align}
Split up each $t_i$ into $t_i=r_i n+\tau_i$ with $\tau_i=t_i\bmod n$ and $r_i \in \mathbb{N}_0$. Then, $A^{t_i} = \gamma^{r_i}A^{\tau_i}$ due to (\ref{eq:eigs}). Hence, if (\ref{eq:set_ex}) is not satisfied, the sum in (\ref{eq:sumA2}) contains at most $n-1$ linearly independent vectors $CA^{\tau_i}$, contradicting the fact that it is nonzero for any choice of $\alpha_i, i=1,\ldots,l$, that are not all zero.
\end{proof}

\subsection{Numerical example for \ref{subsec:hos}}
\label{App:B}
A numerical example is now presented to illustrate the case of a discrete-time system for which no meaningful sampling scheme considering arbitrary measurement instances is possible. Consider a ninth order system with  the following eigenvalues
\begin{align}
	\begin{matrix*}[l]
		&\lambda_1=-0.7500 - 0.2730i&\lambda_2=	-0.7500 + 0.2730i\\ &\lambda_3=-0.3991 - 0.6912i&\lambda_4=-0.3991 + 0.6912i\\
		&\lambda_5=	0.1386 - 0.7860i&	\lambda_6=0.1386 + 0.7860i\\	&\lambda_7=	0.6114 - 0.5130i&	\lambda_8=	0.6114 + 0.5130i\\
		&\lambda_9=	0.7981\\
	\end{matrix*}
\end{align} 
This yields
\begin{align}
	&\phantom{\Rightarrow} \ \  \lambda_p^9=0.1314, \quad p=1, \ldots, 9\\
	&\Rightarrow A^9=0.1314 I \Rightarrow CA^9=0.1314  C
\end{align}
Notice that selecting measurements at the following time instances results in a sample-based observability matrix with a rank less than nine.
\begin{align}
	\begin{matrix}
		&0&1&2&\cdots&7 \\ 
		&9&10&11&\cdots&16\\ 
		&\vdots&\vdots&\vdots&\vdots&\vdots\\
		&9r&9r+1&9r+2&\cdots&9r+7\\ 
	\end{matrix}
\end{align}
These time instances correspond to taking eight out of every nine samples.\\
\subsection{Bound in Remark \ref{rem}}
\label{App:C}
Take $N_s$ samples as in (\ref{eq:boundA2}), at $N_e$ even time instances and $N_o$ odd time instances, with $N_o,N_e \in \mathbb{N}$ and $N_o +N_e=N_s$. Since $N_s \geq (N_p+T)/2$, it follows that $N_e,N_o \geq N_p/2$. Without loss of generality, let $t_1, \ldots, t_{N_e}$ be even and $t_{N_e+1},\ldots, t_{N_s}$ be odd. Define 
\begin{align}
	\mathcal{O}_e=\begin{pmatrix}CA^{t_1}\\ \vdots\\ CA^{t_{N_e}}\end{pmatrix} , \quad  \mathcal{O}_o=\begin{pmatrix}CA^{t_{N_e+1}}\\ \vdots\\ CA^{t_{N_s}}\end{pmatrix}.
\end{align}
Notice that (\ref{eq:obmat_s}) is $\begin{pmatrix}\mathcal{O}_e\\\mathcal{O}_o\end{pmatrix}$. By definition of $N_p$, we have $\lambda_p=-\lambda_q$ for $N_p/2$ pairs of eigenvalues. Therefore, both $\mathcal{O}_e$ and $\mathcal{O}_o$ each have at least $N_p/2$ linearly dependent columns. It can be shown that (compare the case of only positive eigenvalues as in Corollary \ref{cor})
\begin{align}
	\text{rank} \ \mathcal{O}_e=\min\{N_e,n-N_p/2\}, \label{eq:rankOe}\\
		\text{rank} \ \mathcal{O}_o=\min\{N_o,n-N_p/2\}.
	\label{eq:rankOo}
	\end{align}
Now we claim that 
\begin{align}
	\text{rank} \ \begin{pmatrix}\mathcal{O}_e\\\mathcal{O}_o\end{pmatrix}=\min\{n,\text{rank}\ \mathcal{O}_e +\text{rank}\ \mathcal{O}_o\}.
	\label{eq:OeOo}
\end{align}
This can be seen by considering concatenating one row of $\mathcal{O}_o$ to $\mathcal{O}_e$. If the resulting matrix does not have rank greater than that of $\mathcal{O}_e$, it follows that the added row is a linear combination of the rows of $\mathcal{O}_e$ which implies that 
\begin{align}
	\lambda_j^{t_{N_e+1}}=\sum_{i=1}^{N_e}\alpha_i\lambda_j^{t_i}, \quad j=p,q
	\label{eq:sum_eig}
\end{align} 
for some $\alpha_i \in \mathbb{R}$ that are not all zero. Since the additional sampling instance $t_{N_e+1}$ is odd,  \mbox{$\lambda_p^{t_{N_e+1}}=-	\lambda_q^{t_{N_e+1}}$} and, thus, (\ref{eq:sum_eig}) cannot be satisfied  for both $\lambda_p$ and $\lambda_q$. Considering that all other additional $t_i, \ i \geq N_e+2$ are odd and using the above arguments recursively, it can be shown that each additional row increases the rank of the concatenated matrix as long as it is less than $n$, and hence (\ref{eq:OeOo}) holds. Finally, we show that \mbox{$\text{rank}\ \mathcal{O}_e +\text{rank}\ \mathcal{O}_o\geq n$}. Applying (\ref{eq:rankOe}) and (\ref{eq:rankOo}), there are four possible cases
\begin{align}
\begin{aligned}
	&\text{rank}\ \mathcal{O}_e +\text{rank}\ \mathcal{O}_o \\
	&= \begin{cases}
		N_e+N_o=N_s \geq n, \ \text{or} \\
		N_e+n-\frac{N_p}{2} \geq \frac{N_p}{2}+n-\frac{N_p}{2}=n, \ \text{or}\\
		n-\frac{N_p}{2}+N_o \geq n-\frac{N_p}{2}+\frac{N_p}{2}=n, \ \text{or} \\
		n-\frac{N_p}{2}+n-\frac{N_p}{2}=2n-N_p \geq n,
	\end{cases}
\end{aligned}
\end{align}
which shows that $N_s$ samples guarantee sample-based observability.

\begin{thebibliography}{99}
\bibitem{b1} J. W.  Dietrich, G. Landgrafe-Mende, E. Wiora, A. Chatzitomaris, H. H. Klein, J. E. M. Midgley, and R. Hoermann, "Calculated Parameters of Thyroid Homeostasis: Emerging Tools for Differential Diagnosis and Clinical Research", {\it Frontiers in Endocrinology}, vol. 7, June 2016.

\bibitem{b2} V. Brun, A. Eriksen, R. Selseth, K. Johanssson, R. Vik, Renate, B. Davidsen, M. Kaut, and L. Hellemo, "Patient-Tailored Levothyroxine Dosage with Pharmacokinetic/Pharmacodynamic Modeling: A Novel Approach After Total Thyroidectomy", {\it Thyroid}, vol.31, May 2021.

%\bibitem{b10}  	X. Ge, Q.-L. Han, X.-M. Zhang and D. Ding, "Dynamic Event-triggered Control and Estimation: A Survey", {\it International Journal of Automation and Computing}, vol. 18, no. 6, pp.857-886, 2021. 
\bibitem{b10}
M. Miskowicz, "Reducing Communication by Event-Triggered Sampling", {\it Event-Based Control and Signal Processing}, 2015. 
	
\bibitem{b3} 
T. Chen and B. A. Francis, {\it Optimal Sampled-Data Control Systems}, Springer, London, 1995.

\bibitem{b4}
 G. Kreisselmeier, "On sampling without loss of observability/controllability", {\it IEEE Transactions on Automatic Control}, vol. 44, no. 5, pp. 1021–1025, May 1999.
 
\bibitem{b5} 
F. Ding, L. Qiu, and T. Chen, "Reconstruction of continuous-time systems from their non-uniformly sampled discrete-time systems", {\it Automatica}, vol. 45, no. 2, pp. 324-332, 2009.

\bibitem{b6} 
S. Zeng and F. Allgöwer, "A general sampled observability result and its applications", {\it in Proc. 55th IEEE Conference on Decision and Control}, 2016, pp. 3997–4002.

\bibitem{b7} 
L. Y. Wang, C. Li, G. Yin, L. Guo, and C.-Z. Xu, "State observability and observers of linear-time-invariant systems under irregular sampling and sensor limitations", {\it IEEE Transactions on Automatic Control}, vol. 56, no. 11, pp. 2639–2654, 2011.

\bibitem{b8}
C.-T. Chen, {\it Linear System Theory and Design},  Oxford University Press, Oxford, 1999.

\inLongVersion{\bibitem{b85} J.P. Hespanhna, {\it Linear systems theory}, %second.
Princeton University Press, 2018.}

\bibitem{b9} 
B. Anderson, J. Jackson, and M. Sitharam, "Descartes' Rule of Signs Revisited", {\it The American Mathematical Monthly}, vol. 105, no. 5, pp. 447-45, 1998.	
\end{thebibliography}
\end{document}